# Dynamic model of HIV infection with immune system response of T-lymphocytes, B-cells and dendritic cells: a review

**Miguel Ramos-Pascual**

## Abstract

A dynamic model of non-lineal time-dependent ordinary differential equations (ODE) has been applied to the interactions of a HIV infection with the immune system cells. This model has been simplified into two compartments: lymph node and peripheral blood. The model includes CD4 T-lymphocytes in several states (quiescent Q, naive N and activated T), cytotoxic CD8 T-cells, B-cells and dendritic cells. Cytokines and immunoglobulins specific for each antigen (i.e. gp41 or p24) have been also included in the model, modelling the atraction effect of CD4 T-cells to the infected area and the reduction of virus concentration by immunoglobulins. HIV virus infection of CD4 T-lymphocytes is modelled in several stages: before fusion as HIV-attached (H) and after fusion as non-permissive / abortively infected (M), and permissive / latently infected (L) and permissive / actively infected (I). These equations have been implemented in a C++/Python interface application, called Immune System app, which runs Open Modelica software to solve the ODE system through a 4th order Runge-Kutta numerical approximation. Results of the simulation show that although HIV virus concentration in both compartments is lower than $10^{-10}$ virus/$\mu L$ after t=2 years, quiescent lymphocytes reach an equilibrium with a concentration lower than the initial conditions, due to the latency state, which serves as a reservoir in time of virus production. As a conclusion, this model can provide reliable results in other conditions, such as antiviral therapies.

## 1 Introduction

Viral spread is focus of multiple research and modelling in many fields of computerised medicine and biology. Several approaches have been carried out to estimate virus and immune system cells concentrations with time, in which differential equations systems have provided efficient and reliable results [1]. Each model includes in more or less detail the interaction of immune system cells with the virus, simplifying the human body in several compartments (lymphatic system, peripheral blood, neural system or specific tissues, such as skin, epithelium cells, muscles or bones) and several states of the immune cells (quiescent, naive, activated or cytotoxic) [2] [3].

Since 80's Human Immunodeficiency Virus (HIV) has extended worldwide as a viral pandemics which in the worst cases revokes into AIDS (Acquired Immune Deficiency Syndrome), a syndrome characterised by a considerable reduction of CD4 lymphocytes levels and consequently the apparition of secondary oportunistic diseases [4] [5] [6] HIV infection is mainly treated with specific antiviral treatments, as Highly Active Antiretroviral Therapy (HAART), in order to mantain viral loads in undetectable or low concentration levels [7]. No vaccine against HIV has been still developed, although multiple research studies are on-going world-wide [8] [9] .

## 2 Dynamic model of immune system under HIV infection

### 2.1 Introduction

Perelson's model is a set of non-lineal time-dependent ordinary differential equations (ODE) widely used in immunology for modeling in a simplified way the host-cell and virus interactions [1]. Several mathematical models have extended this approach for including for instance cell-to-cell transmissions or infection rates [2]. These models have simplified the reaction of the inmune system, situating the analysis in the worst-case scenario, compared with other research studies [3]

Perelson's model can be extended to multiple compartments, in order to increase the accuracy of results. An extended model in immunology includes two compartments, circulatory and lymphatic system, simplified in two blocks: lymph nodes and peripheral blood. Bone marrow compartment is normally represented with a continuous production rate $\lambda$ of cells

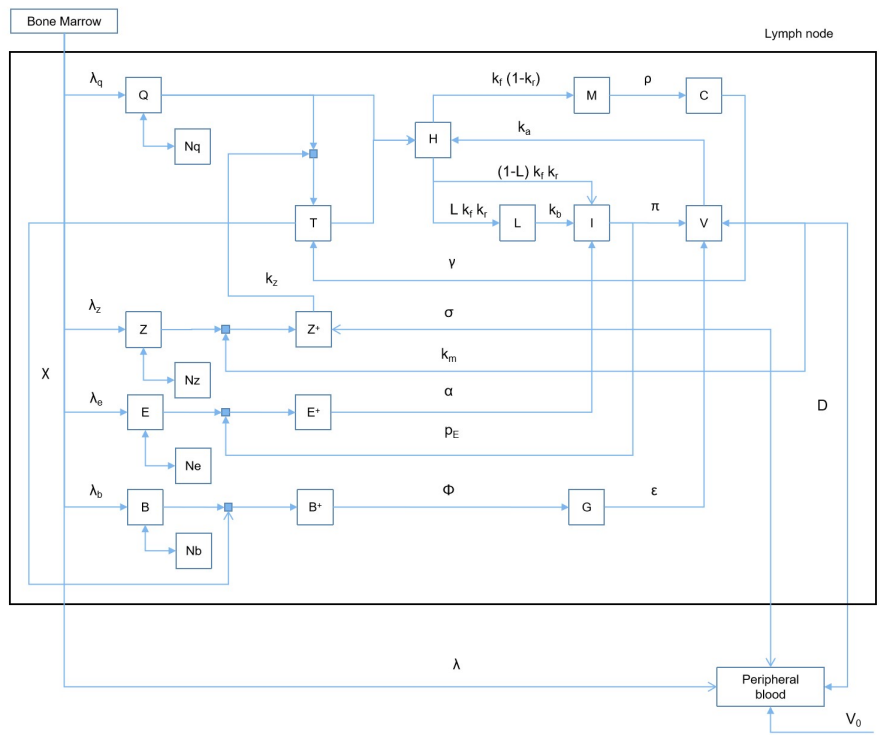

Figure 1: Scheme of Immune System: (1) Bone marrow, (2) lymph node and (3) peripheral blood. Virus infection $V_0$ is produced in peripheral blood compartment

## 2.2 Lymph nodes

### 2.2.1 Introduction

Lymphatic organs are generally divided into primary organs (bone marrow and thymus) and secondary organs (lymph nodes, spleen, tonsil, appendix). In bone marrow, there are pluripotent lymphopoietic stem cells (SCs), which during replication, differentiate into T-cell/B-cell precursors. T-cell precursor cells emigrate to the thymus to maturate into naive CD4 or CD8 T-cells.

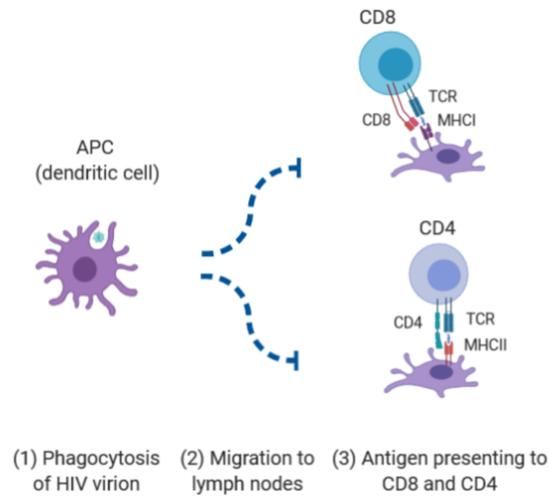

Figure 2: (1) Antigen-presenting-cell (APC) phagocytes a HIV virion and process antigens, (2) migrates to the lymph nodes and (3) presents antigen to T-helper cells (naive CD4 or CD8) through MHC (major-histocompatibility complex), activating CD4 and CD8 T cells (Image produced with https://app.biorender.com)

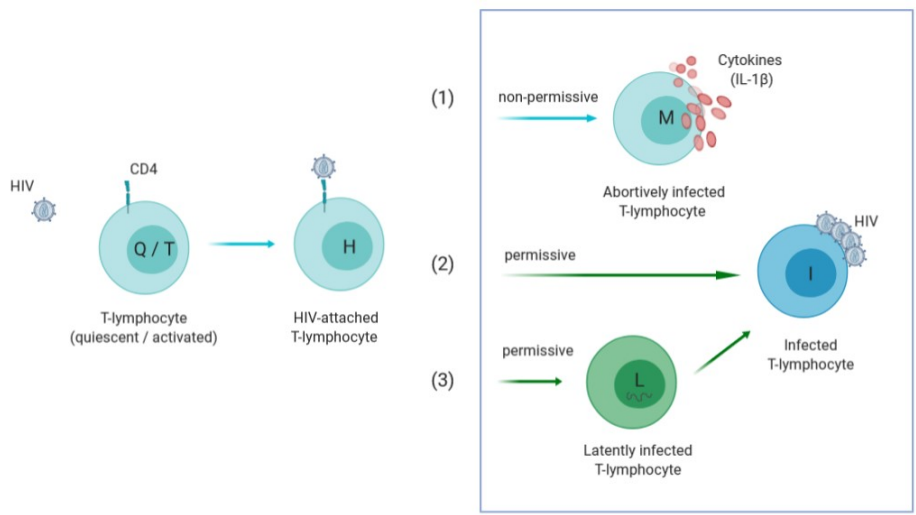

Figure 3: HIV infection of a CD4 T-lymphocyte in three different states: (1) abortively infected (non-permissive) (2) actively infected (permissive) and (3) latently infected (permissive) (Image produced with https://app.biorender.com)

### 2.2.2 CD4 T-lymphocytes

CD4 T-lymphocytes are cycling continuously through the lymphatic system and peripheral blood in a quiescent state $Q$ ($G_0/G_1$ state of virus replication), characterised by low size, low metabolic rates, low levels of transcription and very long periods of survival, until they encounter an antigen-presenting cell (APC), mainly dendritic cells ($Z$). These cells present the antigen to the CD4 T-lymphocytes through the MHCII (major-histocompatibility complex II), activating them ($T$). In a similar way, CD8 T-cells activate into cytotoxic CD8 T-cells when encounter an APC (see fig 2) [10] It is important to remark that T-lymphocytes are activated by APC, that is, activated dendritic cells ($Z^+$) with specific antigens (i.e. gp41, p24, gp120, p17 or p31).

In a first step, HIV virus attaches to the CD4 co-receptor during infection, and then fuses into the cell. This state of lymphocyte before fusion has been called HIV-attached CD4 T-lymphocyte ($H$). After attachment, HIV gp120 spike connects with another co-receptor (CCR5/CXCR4) and then fuses into host-cell. Once virus membrane and capside have merged, reverse-transcription takes place and virus genome is integrated in host-cell nucleous. Provirus remains in a latent state until transcription starts, in a process which is still unknown. This state is called latently infected T-lymphocyte ($L$). Then, once transcription has started, they become actively infected CD4-T lymphocytes ($I$), producing next virus progenie until CD4-T lymphocyte collapses.

In quiescent T-cells, virus replication fails during reverse-transcription because host-cell detection is activated. CD4-T lymphocyte becomes abortively infected or non-permissive to infection ($M$) [11, 10]. Inflammatory cytokines (C) are released from abortively infected cells during pyroptosis, in particular chemokines (IL-1$\beta$), attracting more CD4-T lymphocytes from other areas. Cytokines released during pyroptosis cannot accumulate in blood as in lymph nodes, being unable to attract other immune cells. Therefore, cytokines accumulate only in lymphoid organs [12, 13, 14].

As described in some research studies, HIV virus replication is more permissive in activated than quiescent T-cells, although there are divergent opinions about the reasons. As observed in fig. 3, both processes (permissive and non-permissive infection) are in a certain way similar, because they bring CD4-T lymphocyte to collapse, releasing new virus when infection is permissive or a cytokine storm when is no-permissive. This suggests that HIV virus production and cytokine release can be related processes, or initiated by the same cell subprocess.

In lymph node, equations are expressed for CD4 T-lymphocytes, considering that there is a fraction of circulating CD4 T-cells which are leaving quiescent state for duplicaton ($N_{q1}$), in which infection is not possible, as

$$\frac{\partial Q_1}{\partial t} = \left[ \lambda_{q1} - k_a V_1 Q_1 - k_z Z_1^+ Q_1 - \mu_q Q_1 \right] + \frac{\partial N_{q1}}{\partial t} \qquad (1)$$

$$\frac{\partial T_1}{\partial t} = -k_a V_1 T_1 + k_z Z_1^+ Q_1 + \sigma_{t1}(1 + \gamma C_1) T_2 - \mu_t T_1 \qquad (2)$$

$$\frac{\partial N_{q1}}{\partial t} = r_n Q_1 \left[ 1 - \frac{\sum_j (CD4)_j}{Q_{av1}} \right] - \mu_q N_{q1} \qquad (3)$$

where the sum of all CD4 T-lymphocytes is expressed as

$$\sum_j (CD4)_j = Q_1 + T_1 + N_{q1} + H_1 + L_1 + I_1 + M_1 \qquad (4)$$

where    Q = CD4-T lymphocytes (quiescent)
         T = CD4-T lymphocytes (activated)
         $N_q$ = CD4-T lymphocytes (naive)
         H = CD4-T lymphocytes (HIV-attached)
         L = CD4-T lymphocytes (latently infected)
         I = CD4-T lymphocytes (actively infected)
         M = CD4-T lymphocytes (abortively infected)

and $\lambda_{q1}$ is the production of lymphocytes in the thymus from CD4 T-cell precursors, $k_a$ is the rate of virus attachment to CD4 T-lymphocytes, $k_z$ is the rate of activation of T-lymphocytes from quiescent state after antigen presentation by activated dendritic cells ($Z^+$), $\gamma$ is the efficiency of cytokines on infection, $\mu_q$ is the decay rate of quiescent T cells and $\mu_t$ is the decay rate of activated T cells.

When HIV virus infects a CD4 T-lymphocyte, HIV virion attachs to CD4 co-receptor on host-cell surface as

$$\frac{\partial H_1}{\partial t} = k_a V_1 (Q_1 + T_1) - (k_f + \mu_q) H_1 \qquad (5)$$

where $k_f$ is the rate of virus fusion into CD4 T-cell.

Once the virus has fused into an activated T-lymphocyte and reverse transcription has occurred successfully, CD4 T-lymphocytes stay in a latent state with the provirus integrated in the host-cell genome, until it activates and starts replication, that is

$$\frac{\partial L_1}{\partial t} = L k_f k_r H_1 - (k_b + \mu_t) L_1 \qquad (6)$$

$$\frac{\partial I_1}{\partial t} = (1 - L) k_f k_r H_1 + k_b L_1 - (\alpha E_1^+ + \mu_i) I_1 \qquad (7)$$

where $L$ is the fraction of permissive infections leading to latency, $k_b$ is the activation rate from latent state, $k_r$ is the reverse-transcription and integration rate, $\alpha$ is the killing rate of infected CD4 T-cells by effector or cytotoxic CD8 T-lymphocytes and $\mu_i$ is the decay rate of infected T-cells.

In the case of infection is not permissive and reverse-transcription and integration fail, abortively infected T-lymphocyte and released cytokines are expressed as

$$\frac{\partial M_1}{\partial t} = k_f(1 - k_r)H_1 - \mu_m M_1 \tag{8}$$

$$\frac{\partial C_1}{\partial t} = \rho(\mu_m - \mu_q)M_1 - \mu_{c1}C_1 \tag{9}$$

where $\rho$ is the cytokine production rate per abortively infected cell, $\mu_m$ is the decay rate of abortively infected cells and $\mu_c$ is the decay rate of cytokines.

Virus production is expressed as

$$\frac{\partial V_1}{\partial t} = \pi I_1 - (c + \epsilon G_1)V_1 - k_a V_1(Q_1 + T_1) - k_m V_1 Z_1 - D_1 V_1 + D_2 V_2 \tag{10}$$

where $\pi$ is the virus production rate, $c_1$ is the clearance rate in the lymph node, $\epsilon$ is the efficiency of inmunoglobulins in virus clearance by opsonization and NK cells recruiting and $D_1$ and $D_2$ are the transport rate of virus from blood to lymph node, and viceversa.

### 2.2.3 CD8 T-lymphocytes

CD8-T lymphocytes, which are circulating through lymph nodes and peripheral blood, are activated in the presence of activated and infected lymphocytes into effector CD8+T cells (cytotoxic), with a maximum activation rate of $p_E$, and half-maximal saturation constants of $\theta$ and $\eta$, that is

$$\frac{\partial E_1}{\partial t} = \left[\lambda_{e1} - p_E\left(\frac{I_1}{I_1 + \theta}\right)\left(\frac{T_1}{T_1 + \eta}\right) - \mu_e E_1\right] + \frac{\partial N_{e1}}{\partial t} \tag{11}$$

$$\frac{\partial N_{e1}}{\partial t} = r_e E_1\left[1 - \frac{E_1 + N_{e1} + E_1^+}{E_{av1}}\right] - \mu_e N_{e1} \tag{12}$$

$$\frac{\partial E_1^+}{\partial t} = p_E\left(\frac{I_1}{I_1 + \theta}\right)\left(\frac{T_1}{T_1 + \eta}\right) - \mu_p E_1^+ \tag{13}$$

where $\mu_e$ and $\mu_p$ are the decay rates of CD8+T lymphocytes.

### 2.2.4 Dendritic cells (DC)

In a similar way, activated dendritic cells migrate from the infection place to the lymph nodes to present antigens to the T-cells,

$$\frac{\partial Z_1}{\partial t} = \left[\lambda_{z1} - (k_m V_1 + \mu_z)Z_1\right] + \frac{\partial N_{z1}}{\partial t} \tag{14}$$

$$\frac{\partial N_{z1}}{\partial t} = r_z Z_1\left[1 - \frac{Z_1 + N_{z1} + Z_1^+}{Z_{av1}}\right] - \mu_z Z_1 \tag{15}$$

$$\frac{\partial Z_1^+}{\partial t} = k_m V_1 Z_1 + \sigma_{z1} Z_2^+ - (\sigma_{z2} + \mu_d) Z_1^+ \tag{16}$$

where $k_m$ is the rate of activation and migration of dendritic cells, $\sigma_z$ is the diffusion of DC into peripheral blood and $\mu_z$ is the decay rate of dendritic cells.

### 2.2.5 B-cells

B-cells differentiate in the bone marrow and migrate to the lymphoid organs as inmature B cells, which maturate into naive mature B cells, circulating through the blood and lymph nodes. After the presence of antigen, these B cells develop into plasma B cells, which produce antibodies, or memory B cells, long-lived cells which can be activated in seconday response to antigen.

HIV virus antigens (i.e. glycoproteins, viral envelope/matrix) produce thymus dependent antibody response, requiring helper T cells to synthesize antibodies of more than one isotype (IgM plus IgG, IgA or IgE), what is called a polyclonal response. The interaction of T-helper cells and B cells takes place in the germinal centers of secondary lymphoid organs (as lymph nodes).

In the presence of antigen and a simultaneous signal received from T-helper cell, B cells activate as

$$\frac{\partial B_1}{\partial t} = \left[ \lambda_{b1} - (\xi T_1 + \mu_b) B_1 \right] + \frac{\partial N_{b1}}{\partial t} \tag{17}$$

$$\frac{\partial N_{b1}}{\partial t} = r_b B_1 \left[ 1 - \frac{B_1 + B_1^+ + N_{b1}}{B_{av1}} \right] - \mu_b N_{b1} \tag{18}$$

$$\frac{\partial B_1^+}{\partial t} = \xi T_1 B_1 - \mu_y B_1^+ \tag{19}$$

where $\lambda_{b1}$ is the production rate of B-cells in the lymph node, $\xi$ is the activation rate of B cells for a specific antigen, $\mu_b$ is the decay rate of B cells and $\mu_y$ is the decay rate of plasma B cells.

Once activated, B cells produce immunoglobulins as

$$\frac{\partial G_1}{\partial t} = \phi B_1^+ - (\epsilon V_1 + \mu_g) G_1 \tag{20}$$

where $\phi$ is the production rate of immunoglobulins per activated B cell and $\mu_g$ the decay rate of inmunoglobulins.

A specific response from B-cells is expected from different activated T-lymphocytes by specific antigens or proteins (i.e. gp41, p24, gp120, p17 or p31). For example, immunoglobulins specific for p24 protein will be produced by plasma B-cells activated by activated T-lymphocytes with p24 protein.

### 2.3 Peripheral blood

In peripheral blood compartment, equations are expressed similarly as in lymph node. As an observation, the majority of T-cell lymphocytes circulating through the peripheral blood are in a quiescent state [10]

## 3    Numerical approximation

### 3.1    Iterative solution by Runge-Kutta (4th order)

Let us consider one of the equations of the system of ODE (Ordinary Differential Equations), expressed in integral form as

$$T_1(t) = T_1(t_0) + \int_{t_0}^{t} \big[ -k_a V_1 T_1 + k_z Z_1^+ Q_1 + \sigma_{t1}(1 + \gamma C_1)T_2 - \mu_t T_1 \big] dt \tag{21}$$

and after applying the Runge-Kutta method of 4th order, we arrive to the recursive solution

$$T_{1(n+1)} = T_{1n} + \frac{1}{6}\big[ k_{1n} + 2k_{2n} + 2k_{3n} + k_{4n} \big] \tag{22}$$

where

$$k_{1(n+1)} = h\Big[ -k_a V_{1n} T_{1n} + k_z Z_{1n}^+ Q_{1n} + \sigma_{t1}(1 + \gamma C_{1n})T_{2n} - \mu_t T_{1n} \Big] \tag{23}$$

$$\begin{aligned} k_{2(n+1)} = h\Big[ &-k_a(V_{1n} + k_{1n}/2)(T_{1n} + k_{1n}/2) + k_z(Z_{1n}^+ + k_{1n}/2)(Q_{1n} + k_{1n}/2) \\ &+ \sigma_{t1}(1 + \gamma(C_{1n} + k_{1n}/2))(T_{2n} + k_{1n}/2) - \mu_t(T_{1n} + k_{1n}/2) \Big] \end{aligned} \tag{24}$$

$$\begin{aligned} k_{3(n+1)} = h\Big[ &-k_a(V_{1n} + k_{2n}/2)(T_{1n} + k_{2n}/2) + k_z(Z_{1n}^+ + k_{2n}/2)(Q_{1n} + k_{2n}/2) \\ &+ \sigma_{t1}(1 + \gamma(C_{1n} + k_{2n}/2))(T_{2n} + k_{2n}/2) - \mu_t(T_{1n} + k_{2n}/2) \Big] \end{aligned} \tag{25}$$

$$\begin{aligned} k_{4(n+1)} = h\Big[ &-k_a(V_{1n} + k_{3n})(T_{1n} + k_{3n}) + k_z(Z_{1n}^+ + k_{3n})(Q_{1n} + k_{3n}) \\ &+ \sigma_{t1}(1 + \gamma(C_{1n} + k_{3n}))(T_{2n} + k_{3n}) - \mu_t(T_{1n} + k_{3n}) \Big] \end{aligned} \tag{26}$$

and $h$ is the step height, that is, $t_{n+1} = t_n + h$, and each $k_{in}$ is refered to the function in parenthesis.

In a similar way, we arrive recursively to the solutions $Q_{n+1}$, $H_{n+1}$, $L_{n+1}$, $I_{n+1}$, $M_{n+1}$, $C_{n+1}$, $E_{n+1}$, $P_{n+1}$, $Y_{n+1}$, $Z_{n+1}$, $D_{n+1}$ and $V_{n+1}$.

No time delay has been considered in the simulations, as normally HIV virus cycle from incorporation to next virus progenie release is around 24h in normal conditions [15]

### 3.2    Simulation with Python + Open Modelica (OM)

The dynamic model of HIV infection and immune system response has been modelled and simulated with Python and Open Modelica. A Python script calls to Open Modelica in order to obtain the solutions of the differential equation systems (peripheral blood and lymph node) and then plot them graphically. Open Modelica allows the definition of a series of simulation parameters which define the dynamic model and solve the system of ODE by 4th order Runge-Kutta method.

### 3.3    Simulation parameters

The concentration of CD4-T cells in peripheral blood variates around 200-1500 cells/$\mu L$, depending on the sample and the state of inmuno-depressed system [16, 17]. In a health individual, a normal concentration is around 1000 cells/$\mu$ L. The concentration of CD8-T cells in peripheral blood is around 500 cells/$\mu$ L, which are activated into effector CD8-T cells after antigen-presenting-cells (APC)

In lymph node, CD4-T cell concentration is also variable, as the concentration of CD8-T cells of around 10% in non-infected indiviuals, that is 500 cells/$\mu$ L [18].

Infection rate $k_i$ is calculated as $k_i = k_a k_f k_r$, where $k_a$ is the attachment rate to the CD4 receptor, $k_f$ is the fusion rate with the CCR5/CRCX4 co-receptors and $k_r$ is the reverse-transcription and integration rate.

The initial conditions have been established as the solutions of the steady state system, for the immune system T-cells (CD4 and CD8), B-cells and dendritic cells. The ratio CD4:CD8 has been established as the average value in normal conditions, 3.7 (range from 2 to 11) Table 2 presents the initial conditions of the simulation, with a step height $h = 0.01$ days.

Table 1: Initial conditions of the immune system (t=0)

| Parameter | Lymph node | Blood | Units | Description |
|---|---|---|---|---|
| $Q_0$ | 25000 | 1000 | cells $\mu L^{-1}$ | CD4-T lymphocytes (naive) |
| $E_0$ | 5400 | 270 | cells $\mu L^{-1}$ | CD8-T lymphocytes (naive) |
| $B_0$ | 10000 | 200 | cells $\mu L^{-1}$ | B-cells (naive) |
| $Z_0$ | 1000 | 40 | cells $\mu L^{-1}$ | Dendritic cells (inmature) |
| $T_0$ | 0 | 0 | cells $\mu L^{-1}$ | CD4-T lymphocytes (activated) |
| $H_0$ | 0 | 0 | cells $\mu L^{-1}$ | CD4-T lymphocytes (HIV-attached) |
| $L_0$ | 0 | 0 | cells $\mu L^{-1}$ | CD4-T lymphocytes (latently infected) |
| $I_0$ | 0 | 0 | cells $\mu L^{-1}$ | CD4-T lymphocytes (actively infected) |
| $M_0$ | 0 | 0 | cells $\mu L^{-1}$ | CD4-T lymphocytes (abortively infected) |
| $P_0$ | 0 | 0 | cells $\mu L^{-1}$ | CD8-T lymphocytes (cytotoxic) |
| $Y_0$ | 0 | 0 | cells $\mu L^{-1}$ | B-cells (plasma) |
| $D_0$ | 0 | 0 | cells $\mu L^{-1}$ | Dendritic cells (activated) |
| $C_0$ | 0 | 0 | molecules $\mu L^{-1}$ | Cytokines |
| $G_0$ | 0 | 0 | molecules $\mu L^{-1}$ | Immunoglobulins |

Table 2: HIV concentrations at infection (t=365 days)

| Parameter | Lymph node | Blood | Units | Description |
|---|---|---|---|---|
| $V_0$ | 0 | 1 | virions $\mu L^{-1}$ | HIV virions |

Table 3: Parameters of the immune system used in the simulations

| Parameter | Description | Mean | Units | Reference |
|---|---|---|---|---|
| $Q_{av1}$ | Average population of CD4 T-cell (lymph node) | 50000 | cell $\mu L^{-1}$ | fitted |
| $P_{av1}$ | Average population of CD8 T-cell (lymph node) | 5400 | cell $\mu L^{-1}$ | fitted |
| $B_{av1}$ | Average population of B-cell (lymph node) | 10000 | cell $\mu L^{-1}$ | fitted |
| $Z_{av1}$ | Maximum number of dendritic cells (lymph node) | 1000 | cells $\mu L^{-1}$ | fitted |
| $Q_{av2}$ | Average population of CD4 T-cell (blood) | 1000 | cell $\mu L^{-1}$ | fitted |
| $P_{av2}$ | Average population of CD8 T-cell (blood) | 270 | cell $\mu L^{-1}$ | fitted |
| $B_{av2}$ | Average population of B-cell (blood) | 200 | cell $\mu L^{-1}$ | fitted |
| $Z_{av2}$ | Maximum number of dendritic cells (blood) | 40 | cells $\mu L^{-1}$ | fitted |
| $\lambda_{Q1}$ | CD4 T-cell rate of supply from thymus (lymph node) | 50 | cell $\mu L^{-1}$ day $^{-1}$ | [1] |
| $\lambda_{P1}$ | CD8 T-cell rate of supply from thymus (lymph node) | 55 | cell $\mu L^{-1}$ day $^{-1}$ | [1] |
| $\lambda_{B1}$ | B-cell rate of supply from thymus (lymph node) | 100 | cell $\mu L^{-1}$ day $^{-1}$ | [1] |
| $\lambda_{Z1}$ | Dendritic cell rate of supply from thymus (lymph node) | 10 | cell $\mu L^{-1}$ day $^{-1}$ | [1] |
| $\lambda_{Q2}$ | CD4 T-cell migration rate (blood) | 2 | cell $\mu L^{-1}$ day $^{-1}$ | [1] |
| $\lambda_{P2}$ | CD8 T-cell migration rate (blood) | 2 | cell $\mu L^{-1}$ day $^{-1}$ | [1] |
| $\lambda_{B2}$ | B-cell migration rate (blood) | 2 | cell $\mu L^{-1}$ day $^{-1}$ | [1] |
| $\lambda_{Z2}$ | Dendritic cell migration rate (blood) | 2 | cell $\mu L^{-1}$ day $^{-1}$ | [1] |
| $r_q$ | CD4 T-cell growth rate (quiescent) | 0 | day $^{-1}$ | [1] |
| $r_n$ | CD4 T-cell growth rate (naive) | 0.03 | day $^{-1}$ | fitted |
| $r_e$ | CD8 T-cell growth rate (naive) | 0.03 | day $^{-1}$ | fitted |
| $r_b$ | B-cell growth rate (naive) | 0.03 | day $^{-1}$ | [19] |
| $r_z$ | Dendritic cell growth rate (immature) | 0.03 | day $^{-1}$ | fitted |
| $r_t$ | CD4 T-cell growth rate (activated) | 0.5 | day $^{-1}$ | [1] |
| $r_p$ | CD8 T-cell growth rate (activated) | 0.5 | day $^{-1}$ | fitted |
| $r_y$ | B-cell growth rate (activated) | 0.5 | day $^{-1}$ | [19] |
| $r_d$ | Dendritic cell growth rate (mature) | 0.5 | day $^{-1}$ | fitted |
| $\mu_q$ | Decay rate of quiescent CD4 T-cell | 0.01 | day $^{-1}$ | fitted |
| $\mu_t$ | Decay rate of activated CD4 T-cell | 0.01 | day $^{-1}$ | [13] |
| $\mu_m$ | Decay rate of abortively infected CD4 T-cell | 1 | day $^{-1}$ | [13] |
| $\mu_i$ | Decay rate of actively infected CD4 T-helper cell | 1 | day $^{-1}$ | [13] |
| $\mu_e$ | Decay rate of CD8 T-cell | 0.01 | day $^{-1}$ | [13] |
| $\mu_p$ | Decay rate of CD8 T-cell (cytotoxic) | 0.01 | day $^{-1}$ | [13] |
| $\mu_z$ | Decay rate of dendritic cells | 0.0037 | day $^{-1}$ | [20] |
| $\mu_d$ | Decay rate of dendritic cells (activated) | 0.024 | day $^{-1}$ | [20] |
| $\mu_b$ | Decay rate of B-cells | 0.01 | day $^{-1}$ | [21] |
| $\mu_{c1}$ | Decay rate of cytokine (lymph node) | 2 | day $^{-1}$ | [22] |
| $\mu_{c2}$ | Decay rate of cytokine (peripheral blood) | 10 | day $^{-1}$ | [22] |
| $\mu_g$ | Decay rate of immunoglobulins | 0.03 | day $^{-1}$ | [21] |
| $k_m$ | Activation and migration rate of DC | 1.5e-1 | cell virion $^{-1}$ day $^{-1}$ | [23] |
| $k_z$ | Activation rate of T-cells from quiescent state by DC | 1e-6 | day $^{-1}$ | fitted |
| $\sigma_{z1}$ | Diffusion of DC from blood to lymph node | 1000 | day $^{-1}$ | fitted |
| $\sigma_{z2}$ | Diffusion of DC from lymph node to blood | 1 | day $^{-1}$ | fitted |
| $\rho$ | Cytokine production rate | 15 | molecules cell $^{-1}$ | [13] |
| $\gamma$ | Efficiency of cytokines on infection | 0.2 | $\mu L$ molecule $^{-1}$ | [13] |
| $\alpha$ | Killing rate of CD8+T cells | 10 | $\mu L$ cell $^{-1}$ day $^{-1}$ | [13] |
| $p_E$ | Max activation rate of CD8+T cells | 0.1 | cells $\mu L^{-1}$ day $^{-1}$ | [13] |
| $\theta$ | Half max saturation of CD8+T cells | 0.05 | cells $\mu L^{-1}$ | [13] |
| $\eta$ | Half max saturation of CD4+T cells on CD8+T cell activation | 500 | cells $\mu L^{-1}$ | [13] |
| $\xi$ | Probability of simultaneous activation Tcell/Bcell/antigen | 0.01 | - | fitted |
| $\phi$ | Production rate of antibodies per activated B-cell | 2000 | s $^{-1}$ | [24] |
| $\epsilon$ | Efficiency of Ig in virus clearance by macrophages | 1e-15 | - | fitted |

Table 4: Parameters of the HIV cycle replication used in the simulations

| Parameter | Description | Mean | Units | Reference |
|---|---|---|---|---|
| $k_i$ | Infection rate | 2.4e-5 | $\mu L$ virion$^{-1}$ day$^{-1}$ | [1] |
| $k_a$ | Attachment rate to CD4 T-cell | 5e-3 | $\mu L$ virion$^{-1}$ day$^{-1}$ | fitted |
| $k_f$ | Fusion rate in CD4 T-cell | 1e-1 | $\mu L$ virion$^{-1}$ day$^{-1}$ | fitted |
| $k_r$ | Reverse-transcription and integration rate in CD4 T-cell | 5e-2 | $\mu L$ virion$^{-1}$ day$^{-1}$ | fitted |
| $k_b$ | Activation rate from latently infected state | 5e2 | $\mu L$ virion$^{-1}$ day$^{-1}$ | [1] |
| $L$ | Fraction of permissive infections leading to latency | 5e-2 | - | fitted |
| $\pi$ | Virus burst size per infected cell | 100 | virion cell$^{-1}$ day$^{-1}$ | [25] |
| $c$ | Virus clearance rate | 3 | day$^{-1}$ | [20] |
| $D_1$ | Viral diffusion rate (blood to lymph) | 0.1 | day$^{-1}$ | [13] |
| $D_2$ | Viral diffusion rate (lymph to blood) | 0.2 | day$^{-1}$ | [13] |

### 3.4 Results and discussion

Figure 4 shows the maximum values of immune cells obtained with the immune system application, after an initial infection in the peripheral blood of 1 virion/$\mu L$ at t=1 year.

As observed in figures 5 to 18, immune system response cell concentrations reach an equilibrium after initial infection, with a maximum value at t=0. Although virus concentration is $< 10-10$ virion/$\mu L$ after t>2 years, quiescent T-cells ($Q$) are considerably lower than initial conditions ($Q_1$=25000 cells/$\mu L$ and $Q_2$=1000 cells/$\mu L$), due to the fact that virus is continuously producing new virions from a latency state.

```
-----------------------------------------------------------------------
  IMMUNE SYSTEM (Maximum values) - cells/muL
-----------------------------------------------------------------------
                                   Lymph node   |   Peripheral blood  |
-----------------------------------------------------------------------
  INACTIVATED STATE
-----------------------------------------------------------------------
CD4 T cell (quiescent)          :  Q100 = 2.66E+4    Q200 = 1.00E+3
CD8 T cell (naive)              :  E100 = 5.50E+3    E200 = 2.70E+2
B-cells (naive)                 :  B100 = 1.00E+4    B200 = 2.00E+2
Dendritic cells (immature)      :  Z100 = 2.70E+3    Z200 = 5.40E+2
-----------------------------------------------------------------------
  ACTIVATED STATE
-----------------------------------------------------------------------
CD4 T cell (activated by gp41)  :  T100 = 7.26E-2    T200 = 5.33E-6
CD8 T cell (activated by gp41)  :  P100 = 1.69E-9    P200 = 8.51E-12
Dendritic cells (antigen gp41)  :  D100 = 8.15E-1    D200 = 8.21E-4
B-cells  (plasma antigen gp41)  :  Y100 = 4.28E+2    Y200 = 6.45E-4
Immunoglobulins (antigen gp41)  :  G100 = 2.35E+12   G200 = 3.52E+6
Cytokines                       :  C100 = 1.21E-4    C200 = 2.02E-3
-----------------------------------------------------------------------
  INFECTED STATE
-----------------------------------------------------------------------
HIV virus                       :  V100 = 3.75E-7    V200 = 1.00E+0
-----------------------------------------------------------------------
CD4 T cell (HIV attached)       :  H100 = 1.55E-4    H200 = 1.28E-2
CD4 T cell (abortively infected):  M100 = 1.58E-5    M200 = 1.35E-3
CD4 T cell (latently infected)  :  L100 = 7.75E-11   L200 = 1.28E-7
CD4 T cell (infected)           :  I100 = 8.30E-7    I200 = 7.09E-5
-----------------------------------------------------------------------
```

Figure 4: Immune System app: Maximum values of inactivated, activated and infected states of immune system cells and HIV virus particles

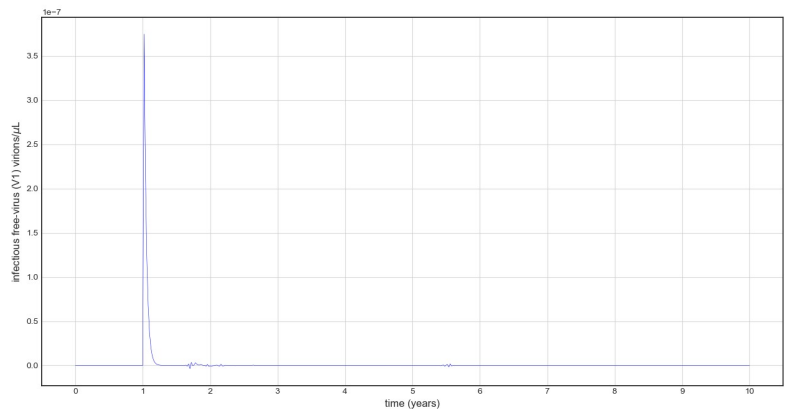

Figure 5: Infectious HIV free-virus concentration ($V_1$) after infection at t=1 year (lymph node)

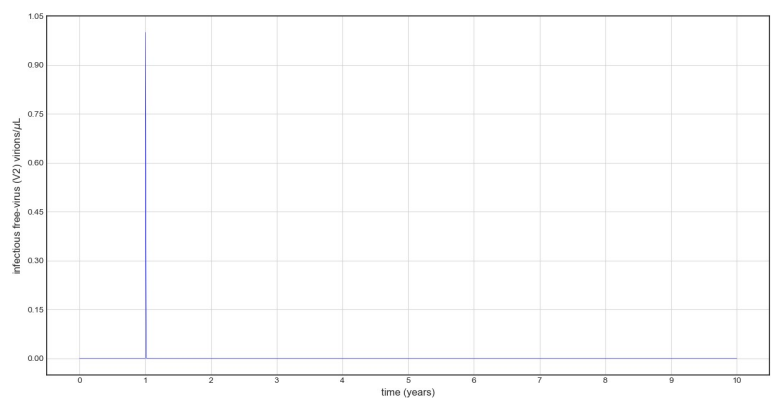

Figure 6: Infectious HIV free-virus concentration ($V_2$) after infection at t=1 year (peripheral blood)

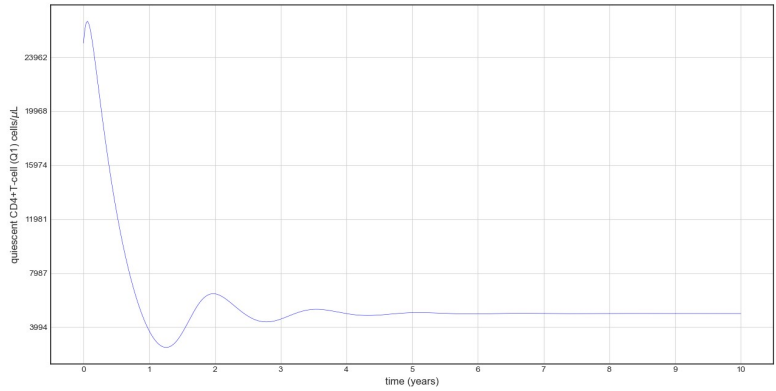

Figure 7: Quiescent CD4 T-lymphocytes ($Q_1$) concentration after infection at t=1 year (lymph node)

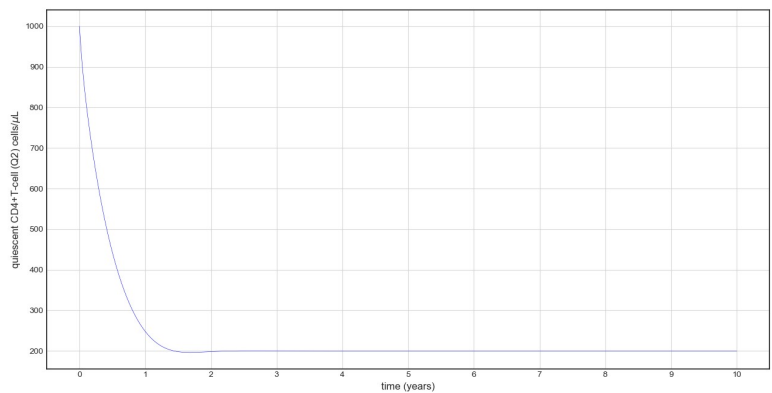

Figure 8: Quiescent CD4 T-lymphocytes ($Q_2$) concentration after infection at t=1 year (peripheral blood)

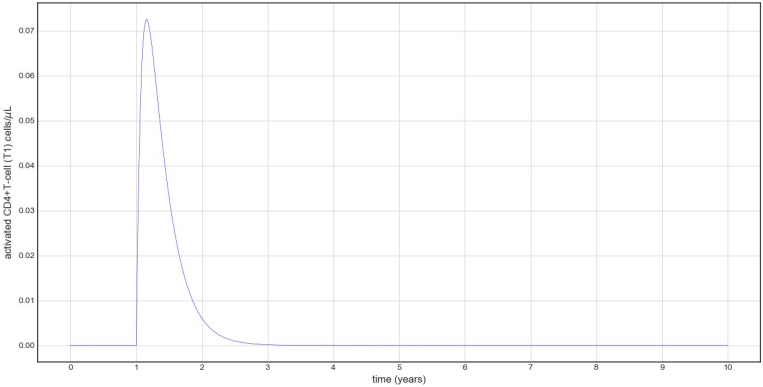

Figure 9: Activated CD4 T-lymphocytes ($T_1$) concentration after infection at t=1 year (lymph node)

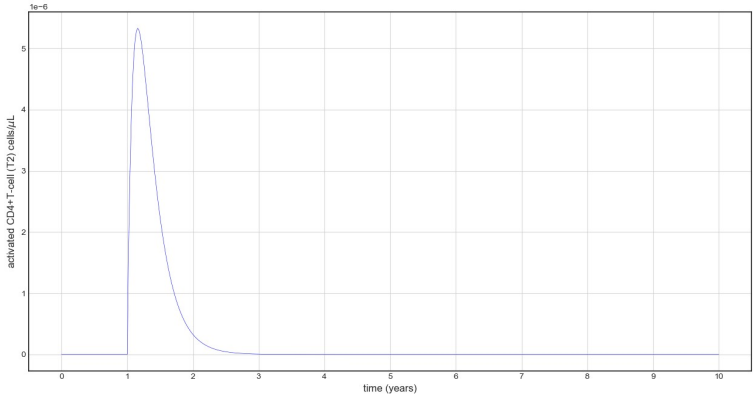

Figure 10: Activated CD4 T-lymphocytes ($T_2$) concentration after infection at t=1 year (peripheral blood)

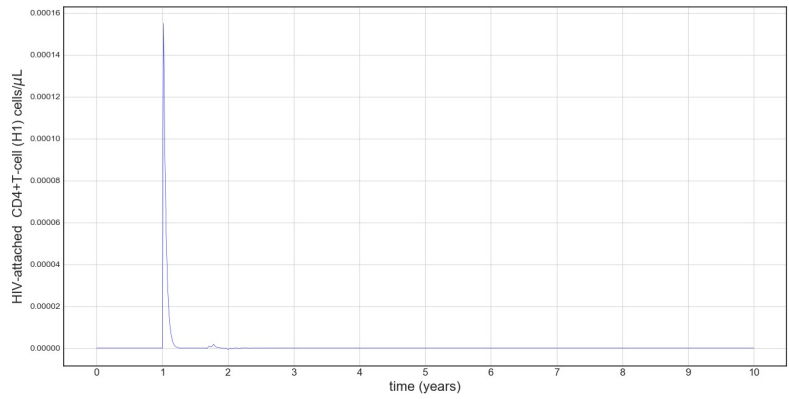

Figure 11: HIV-attached CD4 T-lymphocytes ($H_1$) concentration after infection at t=1 year (lymph node)

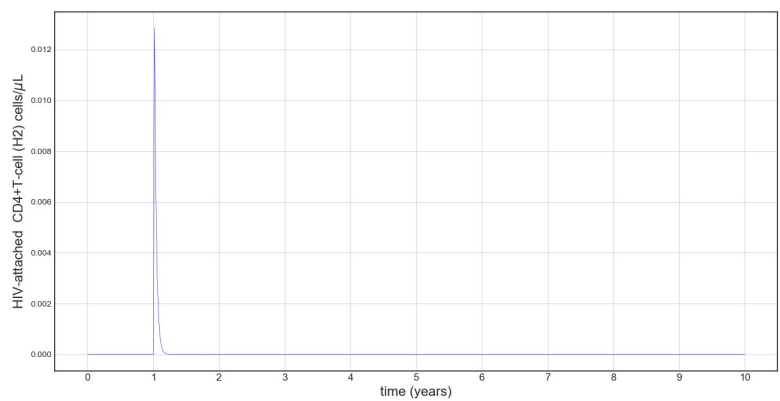

Figure 12: HIV-attached CD4 T-lymphocytes ($H_2$) concentration after infection at t=1 year (peripheral blood)

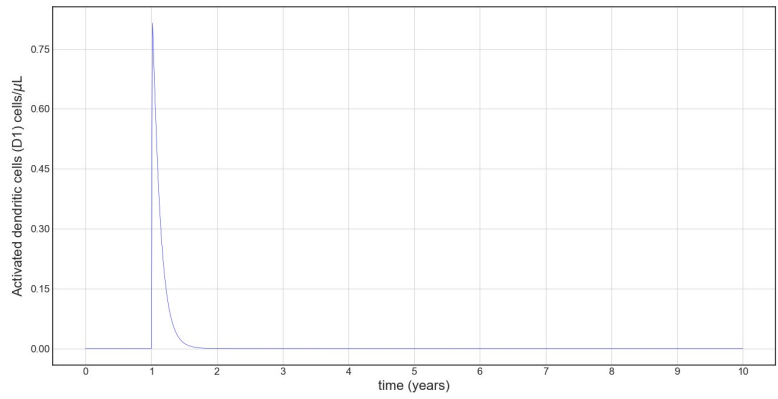

Figure 13: Activated dendritic cells ($Z_1^+$) concentration after infection at t=1 year (lymph node)

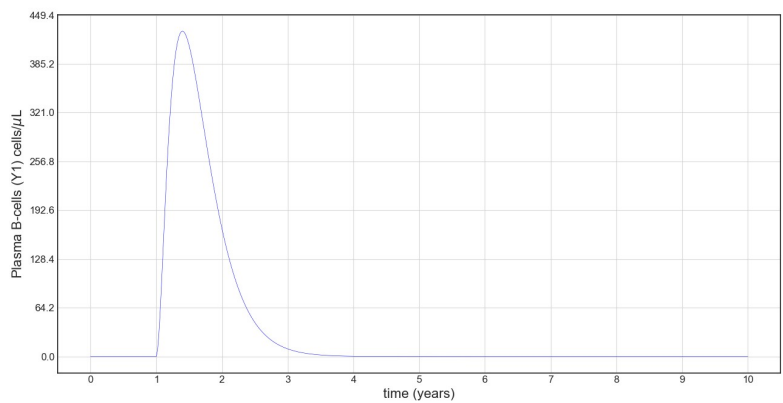

Figure 14: Plasma B-cells ($Y_1$) concentration after infection at t=1 year (lymph node)

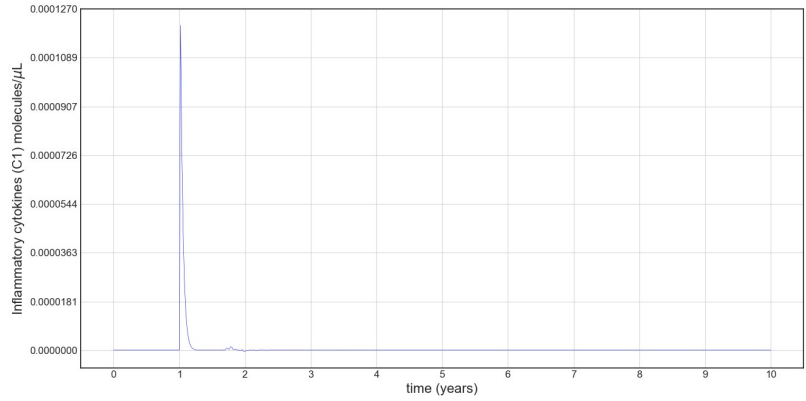

Figure 15: Inflammatory cytokines ($C_1$) concentration after infection at t=1 year (lymph node)

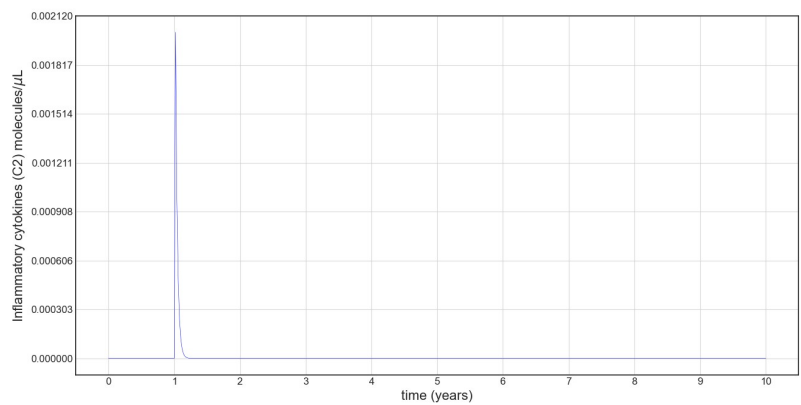

Figure 16: Inflammatory cytokines ($C_2$) concentration after infection at t=1 year (peripheral blood)

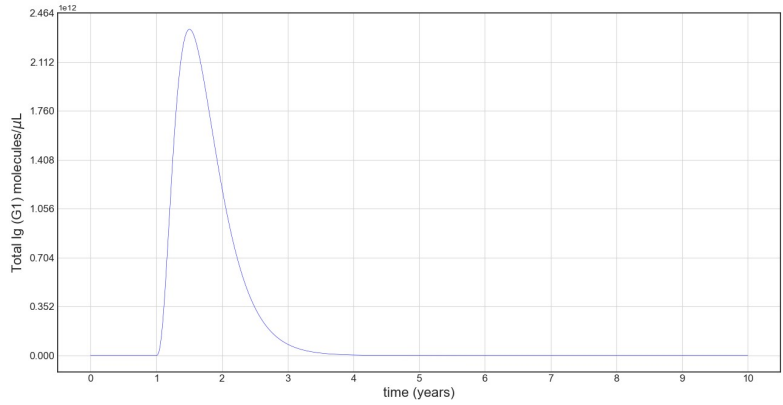

Figure 17: Total immunoglobulins ($G_1$) concentration after infection at t=1 year (lymph node)

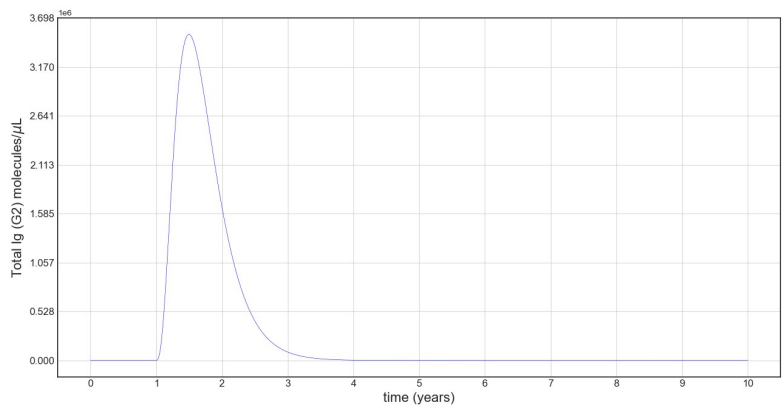

Figure 18: Total immunoglobulins ($G_2$) concentration after infection at t=1 year (peripheral blood)

## 3.5 Conclusions

The dynamic model presents in detail the immune system response against a HIV virus infection in the peripheral blood compartment: activation of dendritic cells (DC), migration to the lymph node, activation of CD4 T-cells, CD8 T-cells and B-cell, with the production of inflammatory cytokines atracting more CD4 T-cells to the infection place and immunoglobulins to attach to the virus particles. The HIV virus infects CD4 T-cells in several stages, from attachment, fusion, reverse-transcription, integration and virus budding. CD4 T-cells can be permissive or non-permissive to infection, resulting in two states: abortively infected T-cells ($M$), latently infected ($L$) and actively infected T-cells ($I$). All these states can be modelled through a system of ordinary differential equations (ODE) which have been solved with numerical methods.

Although the dynamic model provides reliable results with an impulse infection in the peripheral blood (initial conditions), there are several aspects of the modelling still unknown, such as the attachment, fusion and reverse-transcription rates, which has been fitted. Furthermore, activated cells and immunoglobulins have been simulated to only one antigen target (glycoprotein gp41). Further simulations must be applied to other HIV antigens, such as p24, gp120, p17 or p31.

Further research will be focused on stability and endemic equilibrium of the system, focusing on the reproduction number ($R_0$), the disease-free equilibrium (DFE) and the Jacobian matrixes of the system. The model will be applied in the future also to simulate the effect of antiviral drugs against HIV infection, such as the HAART therapy.